\documentclass[9pt,conference]{IEEEtran}
\IEEEoverridecommandlockouts

\usepackage{cite}
\usepackage{amsmath,amssymb,amsfonts}
\usepackage{algorithmic}
\usepackage{graphicx}
\usepackage{textcomp}
\usepackage{xcolor}
\def\BibTeX{{\rm B\kern-.05em{\sc i\kern-.025em b}\kern-.08em
    T\kern-.1667em\lower.7ex\hbox{E}\kern-.125emX}}

\usepackage{arydshln}
\usepackage{bm}
\usepackage{booktabs}
\usepackage{multirow}
\usepackage{hyperref}
\definecolor{darkblue}{rgb}{0, 0, 0.5}
\hypersetup{
    colorlinks=true,
    citecolor=darkblue,
    linkcolor=darkblue,
    urlcolor=darkblue
}

\DeclareMathOperator*{\argmax}{\text{argmax}}

\renewcommand{\paragraph}[1]{\noindent\textbf{#1}\hspace{0.5em}}

\makeatletter
\def\adl@drawiv#1#2#3{%
        \hskip.5\tabcolsep
        \xleaders#3{#2.5\@tempdimb #1{1}#2.5\@tempdimb}%
                #2\z@ plus1fil minus1fil\relax
        \hskip.5\tabcolsep}
\newcommand{\cdashlinelr}[1]{%
  \noalign{\vskip\aboverulesep
           \global\let\@dashdrawstore\adl@draw
           \global\let\adl@draw\adl@drawiv}
  \cdashline{#1}
  \noalign{\global\let\adl@draw\@dashdrawstore
           \vskip\belowrulesep}}
\makeatother

\begin{document}

\title{Harnessing the Zero-Shot Power of Instruction-Tuned Large Language Model for Guiding End-to-End Speech Recognition}

\author
{
\IEEEauthorblockN{
    Yosuke Higuchi,
    Tetsuji Ogawa and
    Tetsunori Kobayashi
}
\IEEEauthorblockA{
    \textit{Department of Communications and Computer Engineering, Waseda University, Tokyo, Japan}
}
}

\maketitle
\setlength{\abovedisplayskip}{4pt}
\setlength{\belowdisplayskip}{4pt}
\setlength{\textfloatsep}{0.4cm} %

\begin{abstract}
We propose to utilize an instruction-tuned large language model (LLM)
for guiding the text generation process in automatic speech recognition (ASR).
Modern LLMs are adept at performing various text generation tasks through zero-shot learning,
prompted with instructions designed for specific objectives.
This paper explores the potential of LLMs to derive linguistic information that
can facilitate text generation in end-to-end ASR models.
Specifically,
we instruct an LLM to correct grammatical errors in an ASR hypothesis and
use the LLM-derived representations to refine the output further.
The proposed model is built on the joint CTC and attention architecture,
with the LLM serving as a front-end feature extractor for the decoder.
The ASR hypothesis, subject to correction, is obtained from the encoder via CTC decoding and
fed into the LLM along with a specific instruction.
The decoder subsequently takes as input the LLM output to perform token predictions,
combining acoustic information from the encoder and the powerful linguistic information provided by the LLM.
Experimental results show that
the proposed LLM-guided model achieves a relative gain of approximately 13\% in word error rates across major benchmarks.
\end{abstract}


\vspace{-0.025cm}
\section{Introduction}
\vspace{-0.025cm}
Driven by rapid advancements in self-supervised pre-training methods~\cite{devlin2019bert,radford2018improving},
large language models (LLMs) have shown remarkable versatility in acquiring various capabilities with minimal or even no task-specific training data~\cite{radford2019language,brown2020language,scao2022bloom,wei2022emergent,chowdhery2023palm}.
This has highlighted the potential of LLMs to address different downstream natural language processing (NLP) tasks in a few-shot or even zero-shot manner,
enabling highly flexible functions that have been beneficial to end users~\cite{openai2023gpt4}.

Such few-/zero-shot learning capabilities of LLMs can be improved by
strategically using the prompting mechanism to guide their behavior.
In-context learning~\cite{brown2020language} allows LLMs to learn tasks without requiring parameter updates,
which is achieved by adding a few input-output examples prior to a target input.
To further enhance the controllability of LLMs,
instruction fine-tuning~\cite{wei2022finetuned,ouyang2022training,chung2024scaling} is employed to make the model adhere more closely to instructions.
This has allowed for generating responses that align with desired outcomes,
substantially boosting zero-shot performance on unseen tasks.

In this work,
we explore the application of the LLM's zero-shot learning capability for solving speech-processing tasks,
focusing on end-to-end automatic speech recognition (ASR).
We aim to align a specific speech-processing task with a related task in NLP and
use the linguistic information embedded within the LLM to facilitate solving the target speech task.
Specifically, we relate the end-to-end ASR task to zero-shot grammatical error correction~\cite{nie2022prompt,wu2023chatgpt,fang2023chatgpt}.
Recent efforts have revealed the effectiveness of using LLMs for ASR error correction as post-processing~\cite{ma2023n,ma2023can,yang2023generative,song2023contextual},
wherein the model is instructed to select the most probable output from an $N$-best hypothesis list.
Our approach, in contrast, attempts to directly integrate the LLM's inherent ability to correct grammatical errors
into an ASR formulation that is optimized in an end-to-end manner.

The proposed model is based on the joint connectionist temporal classification (CTC) and attention architecture~\cite{watanabe2017hybrid}.
The core component of our model is the LLM-guided decoder,
which augments the original decoder by employing a fixed-parameter LLM
to serve as a front-end feature extractor.
This integration allows for enhancing the language modeling capabilities in the decoder,
inheriting the LLM's proficiency in correcting grammatical errors.
Additionally, the cross-attention mechanism facilitates aligning the LLM-derived representations with the speech information embedded by the encoder.
To optimize the extraction of linguistic features beneficial for the decoder,
we also design an effective prompting strategy for the LLM,
using a hypothesized text sequence generated through CTC decoding.

\section{Background}

\subsection{Instruction-Tuned LLM}
The recent LLMs possess the capability to be ``prompted'' to execute specific tasks.
We focus on Llama2-Chat, an instruction-tuned version of Llama2~\cite{touvron2023llamatwo},
as a pre-trained LLM used in this work.
We hereafter refer to this chat model as ``Llama2'' for brevity.
Llama2, consisting of deep Transformer decoder-based layers~\cite{vaswani2017attention,radford2018improving},
outputs a $D^{\mathsf{llm}}$-dimensional hidden vector $\bm{\mathrm{e}}_n$ at a token position $n$ as
\begin{equation}
    \label{eq:e_n}
    \bm{\mathrm{e}}_n = \text{Llama2}(\underbrace{\vphantom{W_{<n}}W^{\mathsf{ins}}}_{\text{Instruction}},\underbrace{\vphantom{W_{<n}}W^{\mathsf{usr}}}_{\text{User Input}},\underbrace{W_{<n}}_{\text{Response}}) \in \mathbb{R}^{D^{\mathsf{llm}}},
\end{equation}
where
$W^{\mathsf{ins}} \in \mathcal{V}^{N^{\mathsf{ins}}}$ is an $N^{\mathsf{ins}}$-length sequence that specifies the instruction of an intended task;
$W^{\mathsf{usr}} \in \mathcal{V}^{N^{\mathsf{usr}}}$ is an $N^{\mathsf{usr}}$-length sequence that serves as the given input for the task;
$W = (w_n \in \mathcal{V} | n=1,\cdots,N)$ is an $N$-length sequence generated by the LLM; and
$\mathcal{V}$ is the vocabulary of Llama2.
The previous tokens are represented as $W_{<n} = (w_0,\cdots,w_{n-1})$,
where $w_0 = \texttt{<s>}$ is a start-of-sentence symbol.
Typically, the combined $W^{\mathsf{ins}}$ and $W^{\mathsf{usr}}$ is referred to as a \textbf{prompt},
which guides the model to generate a response $W$ in a specific manner
(see Fig.~\ref{fig:model} for example).

Llama2 computes the likelihood of a target sequence $W$ as
\begin{equation}
    p(W|W^{\mathsf{ins}},W^{\mathsf{usr}})=\prod_{n=1}^{N + 1} p(w_n|W_{<n},W^{\mathsf{ins}},W^{\mathsf{usr}}), \label{eq:p_llama_W}
\end{equation}
where $w_{N+1} = \texttt{</s>}$ is an end-of-sentence symbol.
The probability of generating $w_n$ in Eq.~\eqref{eq:p_llama_W} is computed
using $\bm{\mathrm{e}}_n$ from Eq.~\eqref{eq:e_n} as
\begin{equation}
    \hspace{-0.1cm}
    p(w_n|W_{<n},W^{\mathsf{ins}},W^{\mathsf{usr}})\!=\!\sigma(\text{Lin}_{D^{\mathsf{llm}} \to |\mathcal{V}|+1}(\bm{\mathrm{e}}_n)) \in [0,1]^{|\mathcal{V}|+1}, \label{eq:p_llama_w}
\end{equation}
where $\text{Lin}_{D^{\mathsf{llm}} \to |\mathcal{V}|+1}(\cdot)$ projects a $D^{\mathsf{llm}}$-dimensional feature 
vector to a logit, and
$\sigma(\cdot)$ represents the softmax function.

\subsection{Joint CTC/Attention-based End-to-End ASR}
\label{ssec:ctcatt}
Let $O \in \mathbb{R}^{T \times F}$ denote a $T$-length input speech sequence
with $F$-dimensional acoustic features and
$W \in \mathcal{V}^{N}$ represent the corresponding target token sequence.
End-to-end ASR aims to directly map $O$ to $W$ by
modeling the posterior distribution of $p(W|O)$ using a single deep neural network.

Attention-based encoder-decoder (AED)~\cite{chorowski2015attention,chan2016listen,vaswani2017attention} formulates end-to-end ASR using a probabilistic chain rule as
\begin{equation}
    p^{\mathsf{aed}}(W|O) \triangleq \prod_{n=1}^{N+1} p(w_n|W_{<n}, O). \label{eq:aed}
\end{equation}
The token emission probability in Eq.~\eqref{eq:aed} is computed as
\begin{align}
    H = (\bm{\mathrm{h}}_1,\cdots,\bm{\mathrm{h}}_{T'}) &= \text{Encoder}(O) \in \mathbb{R}^{T' \times D^{\mathsf{asr}}}, \label{eq:aed_enc} \\
    p(w_n|W_{<n},O) &= \text{Decoder}(W_{<n},H) \in [0,1]^{|\mathcal{V}|+1}, \label{eq:aed_dec}
\end{align}
where $\text{Encoder}(\cdot)$ first down-samples $O$ (i.e., $T' = T/4$) and
then converts it into a sequence of $D^{\mathsf{asr}}$-dimensional hidden vectors $H$.
$\text{Decoder}(\cdot)$ represents autoregressive decoder layers,
followed by a linear layer and the softmax function,
which map to the output vocabulary, $\mathcal{V} \cup \{\texttt{</s>}\}$.
Here, the decoder is equipped with the cross-attention mechanism
for aligning each token in $W_{<n}$ to the encoder output $H$.
The AED model is optimized by minimizing the negative log-likelihood of Eq.~\eqref{eq:aed},
$\mathcal{L}^{\mathsf{aed}} \triangleq - \log p^{\mathsf{aed}} (W|O)$.

CTC~\cite{graves2014towards} formulates end-to-end ASR by evaluating all possible alignments between $O$ and $W$.
To align the sequences at the frame level, $W$ is augmented by allowing repeated occurrences of the same token
and inserting a blank symbol $\texttt{<b>}$.
Let $A = (a_t \in \mathcal{V} \cup \{\texttt{<b>}\}|t=1,\cdots,T')$ be an alignment sequence, and
CTC models the posterior distribution of $p(W|O)$ as
\begin{equation}
    p^{\mathsf{ctc}}(W|O) \triangleq \sum_{A \in \mathcal{B}^{-1}(W)} \prod_{t=1}^{T'}p(a_t|O), \label{eq:ctc}
\end{equation}
where $\mathcal{B}:A \mapsto W$ is a collapsing function that removes repeated tokens and blank symbols in $A$, and
$\mathcal{B}^{-1}(W)$ is a set of all possible alignments compatible with $W$.
Using the encoder output $H$ from Eq.~\eqref{eq:aed_enc},
the token emission probability in Eq.~\eqref{eq:ctc} is computed as
\begin{equation}
    p(a_t|O) = \text{CTC}(\bm{\mathrm{h}}_t) \in [0,1]^{|\mathcal{V}|+1}, \label{eq:a_t_O}
\end{equation}
where $\text{CTC}(\cdot)$ represents a linear layer and the softmax function
that maps to the output vocabulary of CTC,
$\mathcal{V}\cup\{\texttt{<b>}\}$.
The CTC model is optimized by minimizing the negative log-likelihood of Eq.~\eqref{eq:ctc},
$\mathcal{L}^{\mathsf{ctc}} \triangleq - \log p^{\mathsf{ctc}} (W|O)$.

AED and CTC can be effectively combined to enhance robustness during training and inference processes of end-to-end ASR~\cite{kim2017joint,watanabe2017hybrid}.
The objective function of the joint model is defined as a linear interpolation of $\mathcal{L}^{\mathsf{ctc}}$ and $\mathcal{L}^{\mathsf{aed}}$ as
$\mathcal{L}^{\mathsf{ctc}\text{-}\mathsf{aed}} = \lambda \mathcal{L}^{\mathsf{ctc}} + (1 - \lambda)\mathcal{L}^{\mathsf{aed}}$,
where $\lambda$ ($0 < \lambda < 1$) is a tunable weight.
Joint decoding is performed using a one-pass beam search,
with the score of a hypothesis $\hat{W}$ calculated using Eqs.~\eqref{eq:aed} and~\eqref{eq:ctc} as
$\xi \log p^{\mathsf{ctc}}(\hat{W}|O) + (1 - \xi) \log p^{\mathsf{aed}} (\hat{W}|O)$,
where $\xi$ ($0 \le \xi \le 1$) is a tunable weight to define the importance of the CTC score.
See~\cite{hori2017joint} for a detailed decoding algorithm.

\section{End-to-End ASR Guided by Instruction-Tuned LLM}
\label{sec:proposed}

\begin{figure}[t]
    \centering
    \vspace{-0.1cm}
    \includegraphics[width=0.95\linewidth]{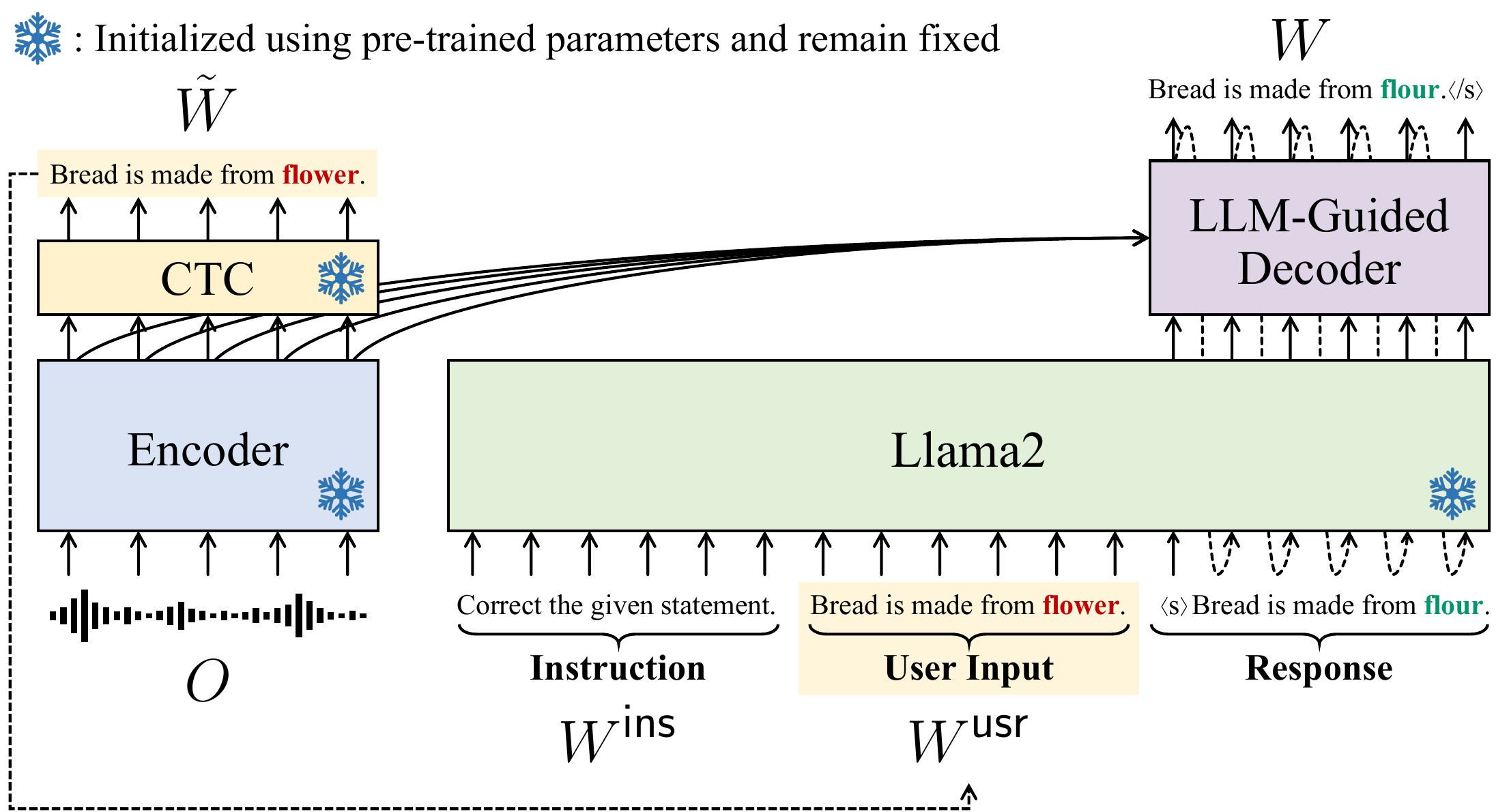} \\
    \vspace{-0.1cm}
    \caption{Proposed end-to-end ASR model guided by instruction-tuned LLM.}
    \label{fig:model}
\end{figure}

\paragraph{Overview} Figure~\ref{fig:model} illustrates the proposed LLM-guided end-to-end ASR model.
Our model is built on the joint CTC and attention framework (as formulated in Sec.~\ref{ssec:ctcatt}),
modifying the decoder by introducing an instruction-tuned LLM as its frontend,
which we refer to as \textbf{LLM-guided decoder}.
Here, the LLM is specifically employed to enhance the language modeling capabilities of the decoder,
aiming to improve ASR performance.
To optimally derive linguistic features that facilitate the text generation process in the decoder,
we capitalize on the LLM's potential as a zero-shot grammatical error correction model~\cite{wu2023chatgpt,fang2023chatgpt,yang2023generative},
designing an effective prompting strategy.
Specifically, a hypothesized output sequence from CTC is
fed into the LLM along with explicit instruction to correct word errors.
The decoder then uses the LLM-derived features to refine the output further,
while the cross-attention mechanism incorporates speech information from the encoder
to mitigate hallucinations and overcorrections.

The following section delves into a precise formulation that substantiates the effectiveness of our model design,
which is followed by detailed descriptions of training and inference strategies.

\subsection{Formulation}
The proposed model formulates end-to-end ASR by
factorizing the posterior distribution $p(W|O)$ as
\begin{equation}
    \label{eq:p_W_O}
    p(W|O) = \sum_{\tilde{W} \in \mathcal{H}(W)} p(W | \tilde{W} ,O) p(\tilde{W} | O),
\end{equation}
where $\tilde{W} \in \mathcal{V}^{M}$ is an $M$-length hypothesized output sequence, and
$\mathcal{H}(W)$ represents a set of all possible output sequences compatible with $W$.
In other words,
$\mathcal{H}(W)$ comprises sequences that are prone to be misrecognized from input speech $O$,
with $\tilde{W}$ derived from $p(\tilde{W}|O)$.
In Eq.~\eqref{eq:p_W_O},
we further factorize $p(W | \tilde{W} ,O)$ by applying a probabilistic chain rule as
\begin{equation}
    \label{eq:p_W_hW_O}
    p(W|\tilde{W},O) = \prod_{n=1}^{N+1} p(w_n|\tilde{W},W_{<n},O).
\end{equation}
Eq.~\eqref{eq:p_W_hW_O} follows the same formulation as AED in Eq.~\eqref{eq:aed},
but it additionally conditions the hypothesized output $\tilde{W}$.
More intuitively,
Eq.~\eqref{eq:p_W_hW_O} can be interpreted as a model
that estimates each current token based on previously predicted tokens $W_{<n}$,
while also \textit{recovering errors present in the hypothesized output sequence} $\tilde{W}$.

The token emission probability in Eq.~\eqref{eq:p_W_hW_O} is modeled similarly to the AED model,
with a modification to the decoder (i.e., Eq.~\eqref{eq:aed_dec}) as
\begin{align}
    &\hspace{-0.117cm}\bm{\mathrm{e}}_n  = \text{Llama2}(W^{\mathsf{ins}}, \tilde{W}, W_{<n}) \in \mathbb{R}^{D^{\mathsf{llm}}}, \label{eq:llama2}\\
    &\hspace{-0.117cm}p(w_n|\tilde{W},W_{<n},O) \nonumber \\
    &\ \ = \text{LLMGuidedDecoder}(\bm{\mathrm{e}}_1,\cdots,\bm{\mathrm{e}}_{n}, H) \in [0,1]^{|\mathcal{V}| + 1}, \label{eq:trfdec}
\end{align}
where $H$ is the encoder output, as derived from Eq.~\eqref{eq:aed_enc}.
In Eq.~\eqref{eq:llama2}, the Llama2 output $\bm{\mathrm{e}}_{n}$ is obtained as in Eq.~\eqref{eq:e_n},
where a hypothesized output sequence is used as the user input, i.e., $W^{\mathsf{usr}}=\tilde{W}$,
accompanied by an instruction $W^{\mathsf{ins}}$ that directs the LLM toward the grammatical error correction task
(see Sec.~\ref{ssec:experimental_setup} for the actual prompt).
Such a prompting strategy is expected to facilitate the modeling of Eq.~\eqref{eq:p_W_hW_O}.
Eq.~\eqref{eq:trfdec} represents the LLM-guided decoder,
the key component of the proposed model,
which is identical to the standard decoder in Eq.~\eqref{eq:aed_dec}
but takes the Llama2 outputs $(\bm{\mathrm{e}}_1,\cdots,\bm{\mathrm{e}}_{n})$ as input.

\subsection{Inference}
\label{ssec:inference}
The most probable sequence is estimated by solving Eq.~\eqref{eq:p_W_O} as
\begin{align}
    \hspace{-0.15cm}\hat{W}\!&= \argmax_{W}\!\sum_{\tilde{W} \in \mathcal{H}(W)}\!p(W | \tilde{W} ,O) p(\tilde{W} | O),\! \label{eq:dec} \\
    &\approx \argmax_{W} p(W | \tilde{W}', O),\ 
    \text{where}\ \tilde{W}'  = \argmax_{\tilde{W}} p(\tilde{W}|O). \label{eq:dec_ctc}
\end{align}
To handle the intractable summation over $\tilde{W}$ in Eq.~\eqref{eq:dec},
we apply the Viterbi approximation with respect to $p(\tilde{W}|O)$,
which results in Eq~\eqref{eq:dec_ctc}.
In Eq.~\eqref{eq:dec_ctc}, $\tilde{W}'$ is first obtained through
the best path decoding of CTC~\cite{graves2006connectionist} using probabilities calculated by Eq.~\eqref{eq:a_t_O},
which involves finding the most probable alignment $\hat{A}$ by concatenating the most active tokens at each time frame,
$\hat{a}_t = \argmax_{a_t} p(a_t|O)$, and
applying the collapsing function as $\tilde{W}'=\mathcal{B}(\hat{A})$.
We then use this CTC output $\tilde{W}'$ to compute $p(W | \tilde{W}', O)$ as in Eq.~\eqref{eq:p_W_hW_O},
which is combined with the CTC score derived from Eq.~\eqref{eq:ctc} to perform the joint CTC and attention decoding~\cite{hori2017joint}.

\subsection{Training}
\label{ssec:training}
The training process for the proposed model unfolds in two steps:
\begin{enumerate}
    \item Train a joint CTC and attention model using $\mathcal{L}^{\mathsf{ctc}\text{-}\mathsf{aed}}$.
    \item Using $\text{Encoder}(\cdot)$ and $\text{CTC}(\cdot)$ trained in Step 1 and pre-trained $\text{Llama2}(\cdot)$, train only $\text{LLMGuidedDecoder}(\cdot)$ (in Eq.~\eqref{eq:trfdec}).
\end{enumerate}
We freeze all the pre-trained networks in Step 2 not only to enhance training efficiency
but also to enable the LLM-guided decoder to focus exclusively on text generation,
ensuring to align the LLM-derived features with the speech information from the encoder.

In Step 2,
the objective function of the proposed model is defined by the negative log-likelihood of Eq.~\eqref{eq:p_W_O} expanded with Eq.~\eqref{eq:p_W_hW_O},
\begin{align}
    - & \log \sum_{\tilde{W} \in \mathcal{H}(W)} \prod_{n=1}^{N+1} p(w_n|\tilde{W},W_{<n},O) p(\tilde{W} | O) \label{eq:loss1} \\
    &\leq \underbrace{-\mathbb{E}_{\tilde{W} \sim p(\tilde{W}|O)} \left[ \log \prod_{n=1}^{N+1} p(w_n|\tilde{W},W_{<n},O) \right]}_{\triangleq \mathcal{L}^{\mathsf{llm}\text{-}\mathsf{dec}}}. \label{eq:loss}
\end{align}
The intractable marginalization over $\tilde{W}$ in Eq.~\eqref{eq:loss1} is
approximated by computing the expectation with respect to the sampling distribution
based on the probability distribution of $p(\tilde{W}|O)$,
which results in Eq.~\eqref{eq:loss}.
Practically, $\mathcal{L}^{\mathsf{llm}\text{-}\mathsf{dec}}$ is computed in a manner similar to the AED loss $\mathcal{L}^{\mathsf{aed}}$,
calculating the cross-entropy losses at each token prediction.
The sampling process of $\tilde{W}$ in Eq.~\eqref{eq:loss} is implemented by
running the encoder in ``training mode'' (with dropout enabled) and performing the best path decoding of CTC,
which is a similar strategy commonly utilized in uncertainty estimation~\cite{gal2016dropout,vyas2019analyzing}.

\section{Related Work}
\paragraph{LLMs with Speech Input}
Prior studies have explored
the incorporation of speech information in LLMs~\cite{wang2023viola,deshmukh2024pengi},
primarily focusing on enabling LLMs to accept speech input.
To convert speech into the input space of LLMs,
discrete representations are typically derived from pre-trained acoustic models~\cite{zhang2023speechgpt,rubenstein2023audiopalm}.
Alternatively, pre-trained ASR models can be used to compress speech into a more manageable length~\cite{chen2023x,nachmani2024spoken,wu2023decoder,yu2024connecting,fathullah2024prompting,baskar2024speech}.
Our approach to utilizing LLMs for speech tasks is related to these studies,
but differs conceptually in that we do not seek to directly adapt LLMs to speech.
Instead, the proposed model aims to extract linguistic features from LLMs for improving ASR models,
without requiring fine-tuning or modifications to off-the-shelf LLMs.

\paragraph{LM Integration in End-to-End ASR}
It has been a widely adopted practice to use separate LMs to improve the performance of end-to-end ASR systems.
The traditional approach includes rescoring~\cite{mikolov2010recurrent,chan2016listen}.
Advanced methods include incorporating LMs into beam search or directly into the model architectures,
through techniques like shallow fusion~\cite{hannun2014deep,gulcehre2015using} and deep fusion~\cite{gulcehre2015using,sriram2018cold,shan2019component}.
More recently, these LM integration approaches have been promisingly extended for use with LLMs~\cite{udagawa2022effect,hu2023massively,chen2024it}.
While the above methods primarily focus on enhancing ASR models with LM information at the output level,
our approach utilizes an LLM as a feature extractor for the decoder to guide its text generation.
Nonetheless, we show that the proposed model remains compatible with conventional LM fusion techniques.

\paragraph{Two-Pass End-to-End ASR}
In ASR, it is a common practice to employ a second-pass model to refine outputs produced by a first-pass model~\cite{sundermeyer2015feedforward,chan2016listen,kannan2018analysis,salazar2020masked}.
End-to-end modeling of ASR has allowed to jointly train
the first-pass and second-pass models~\cite{xia2017deliberation,sainath2019two,hu2020deliberation,wang2022deliberation,higuchi2022bert,higuchi2023mask}.
The proposed formulation in Eq.~\eqref{eq:p_W_O}
shares similarities with these two-pass approaches.
However, it differs in that
the LLM-guided decoder does not specifically deliberate on hypotheses to generate a text sequence.
Instead, it leverages a hypothesized output to derive linguistic features from the LLM.

\begin{table*}[t]
    \centering

    \caption{WERs [\%] ($\downarrow$) and RTF ($\uparrow$) of our models featuring LLM-guided decoder, compared to joint CTC/attention baselines. $B$ and $\xi$ denote the beam size and CTC score weight, respectively, used for the joint CTC/attention decoding.}
    \label{tb:main_results}
    \vspace{-0.25cm}
    \renewcommand{\arraystretch}{0.95}
    \scalebox{0.95}{
    \begin{tabular}{lccccccccccccccc}
        \toprule
        & & & \multicolumn{5}{c}{\textbf{LibriSpeech-100h}} & \multicolumn{4}{c}{\textbf{LibriSpeech-960h}} & \multicolumn{2}{c}{\textbf{TED-LIUM2}} & \multicolumn{2}{c}{\textbf{CoVoST2}} \\
        \cmidrule(l{0.3em}r{0.3em}){4-8}\cmidrule(l{0.3em}r{0.3em}){9-12}\cmidrule(l{0.3em}r{0.3em}){13-14}\cmidrule(l{0.3em}r{0.3em}){15-16}
        & & & \multicolumn{2}{c}{Dev WER} & \multicolumn{2}{c}{Test WER} & \multirow{2}{*}[-3.2pt]{\shortstack[c]{Test\\RTF}} & \multicolumn{2}{c}{Dev WER} & \multicolumn{2}{c}{Test WER} & \multirow{2}{*}[-3.2pt]{\shortstack[c]{Dev\\WER}} & \multirow{2}{*}[-3.2pt]{\shortstack[c]{Test\\WER}} & \multirow{2}{*}[-3.2pt]{\shortstack[c]{Dev\\WER}} & \multirow{2}{*}[-3.2pt]{\shortstack[c]{Test\\WER}} \\
        \cmidrule(l{0.3em}r{0.3em}){4-5}\cmidrule(l{0.3em}r{0.3em}){6-7}
        \cmidrule(l{0.3em}r{0.3em}){9-10}\cmidrule(l{0.3em}r{0.3em}){11-12}
        \texttt{ID}\ \ \textbf{Model} & $B$ & $\xi$ & clean & other & clean & other & & clean & other & clean & other \\
        \midrule
        \texttt{A0}\ \ Joint CTC/Attention & 1 & 1.0 & 9.3 & 21.6 & 9.7 & 22.2 & 0.106 & 3.5 & 8.8 & 3.8 & 8.8 & 10.0 & 9.3 & 23.7 & 26.3 \\
        \cmidrule(l{0.3em}r{0.3em}){1-16}
        \texttt{A1}\ \ Joint CTC/Attention & 1 & 0.3 & 9.9 & 20.6 & 10.7 & 21.1 & 0.147 & 3.4 & 7.4 & 3.6 & 7.5 & 11.6 & 8.8 & 18.9 & 21.8 \\
        \texttt{A2}\ \ + LLM-Guided Decoder & 1 & 0.3 & \textbf{6.7} & \textbf{17.5} & \textbf{7.3} & \textbf{17.9} & 0.222 & \textbf{2.4} & \textbf{6.2} & \textbf{2.6} & \textbf{6.4} & \textbf{9.5} & \textbf{7.8} & \textbf{15.6} & \textbf{18.1} \\
        \cmidrule(l{0.3em}r{0.3em}){1-16}
        \texttt{A3}\ \ Joint CTC/Attention & 20 & 0.3 & 7.2 & 17.5 & 7.5 & 18.0 & 0.194 & 2.7 & 6.6 & 2.9 & 6.7 & 9.4 & 7.8 & 16.2 & 18.4 \\
        \texttt{A4}\ \ + LLM-Guided Decoder & 20 & 0.3 & \textbf{6.2} & \textbf{16.5} & \textbf{6.7} & \textbf{16.9} & 1.193 &  \textbf{2.4} & \textbf{6.2} & \textbf{2.5} & \textbf{6.2} & \textbf{7.6} & \textbf{7.2} & \textbf{15.0} & \textbf{16.9} \\
        \bottomrule
    \end{tabular}
    }
    \vspace{-0.3cm}
\end{table*}

\section{Experiments}
\vspace{-0.05cm}
\label{sec:experiments}

\subsection{Expeirmental Setup}
\label{ssec:experimental_setup}
We used ESPnet~\cite{watanabe2018espnet} for conducting our experiments, and
all the codes are made publicly available at \url{https://github.com/YosukeHiguchi/espnet/tree/llm-guided-asr}.

\paragraph{Data}
We focused on English ASR tasks,
using various corpora spanning different amounts of data and domains,
including LibriSpeech (LS)~\cite{panayotov2015librispeech},
TED-LIUM2 (TED2)~\cite{rousseau2014enhancing}, and
CoVoST2 (CV2)~\cite{wang2021covost}.
In addition to the full 960-hour training set in LS (LS-960),
we used the \textit{train-clean-100} set (LS-100) to explore a lower-resource scenario and
conduct further investigations.
We used CV2 for training models with punctuation and casing preserved,
which can be crucial for the LLM to accurately process text sequences.
During the evaluation, we removed punctuations to exclusively assess ASR performance.

\paragraph{Modeling}
We developed our baseline models within the joint CTC/attention framework (see Sec.~\ref{ssec:ctcatt}),
which used the Conformer-based architecture~\cite{gulati2020conformer,guo2021recent}.
This baseline model also corresponds to the initial joint CTC/attention model trained in Step 1 in Sec.~\ref{ssec:training}.
The proposed model was constructed by substituting the decoder of the baseline model
with the LLM-guided decoder (as detailed in Step 2 in Sec.~\ref{ssec:training}).
The encoder consisted of two convolutional neural network layers followed by a stack of $12$ Conformer blocks.
The number of head $D^{\mathsf{head}}$,
the dimension of a self-attention layer $D^{\mathsf{asr}}$,
the dimension of a feed-forward network $D^{\mathsf{ff}}$, and
the kernel size were set to
($4$, $256$, $1024$, $31$).
The decoder was a stack of $6$ Transformer decoder blocks,
with ($D^{\mathsf{head}}$, $D^{\mathsf{asr}}$, $D^{\mathsf{ff}}$) set to
($4$, $256$, $2048$).
For the LLM, we used Llama2-Chat (7B) provided by HuggingFace~\cite{transformers2020wolf}.
To connect Llama2 with the LLM-guided decoder,
we applied a linear layer to map the dimensions from $D^{\mathsf{llm}}=4096$ to $D^{\mathsf{asr}}=256$.

\paragraph{Training and Decoding}
We primarily followed optimization configurations specified in the ESPnet recipes for each dataset.
The baseline model was trained for up to 50 epochs, and subsequently,
the proposed model featuring the LLM-guided decoder was trained for up to 50 epochs for LS-100
and up to 25 epochs for the other datasets.
For both the baseline and proposed models,
we set the score weight $\xi$ to $0.3$ during the joint CTC/attention decoding,
unless specified otherwise.
The beam size $B$ was set to either $1$ or $20$.

\paragraph{Prompting}
We heuristically designed a prompt to guide Llama2 in performing grammatical error correction,
setting $W^{\mathsf{ins}}$ (in Eq.~\eqref{eq:llama2}) to ``\textit{You will be provided with a statement in quotes. Correct the wrong words and provide your revised version.}''.
As specified in this instruction,
we enclosed a user input (i.e., a hypothesis) within double quotation marks,
which has been found crucial for the model to accurately identify the target sequence.
Through qualitative observation,
we selected a prompt that adhered accurately to the specified task and minimized hallucinations.
Additionally, we aimed to keep the prompt concise to reduce memory usage during training.

\subsection{Main Results}
\vspace{-0.05cm}
\label{ssec:main_results}
Table~\ref{tb:main_results} presents a comparison of the baseline and proposed models across all the tasks,
evaluated by the word error rate (WER).
The \texttt{A0} results represent the CTC decoding performance, with $\xi$ set to $1.0$.
Setting $\xi$ to $0.3$ for the joint decoding resulted in improvements across the tasks (\texttt{A0} vs.\ \texttt{A1}).
The proposed model, featuring the LLM-guided decoder, achieved the best overall performance among the results with a beam size of $B=1$ (\texttt{A0}, \texttt{A1} vs.\ \texttt{A2}).
This indicates the successful incorporation of the LLM into the text generation process,
while only requiring the retraining of minimal parameters (i.e., $18.8$M).
In the proposed inference algorithm (detailed in Sec.~\ref{ssec:inference}),
the CTC decoding results (\texttt{A0}) served as an input to the LLM (i.e., $\tilde{W}$ in Eq.~\eqref{eq:llama2}).
Thus, the gains from \texttt{A0} suggest that the proposed model effectively used the LLM to improve the hypothesized outputs, as expected in Eq.~\eqref{eq:p_W_hW_O}.
In CV2, the proposed model demonstrated a notably higher level of improvements,
particularly considering the gains from the CTC decoding results in \texttt{A0}.
We attribute this to the use of unnormalized text,
which enabled the LLM to extract precise linguistic information.
When we set $B$ to $20$, the proposed model consistently outperformed the baseline across all tasks (\texttt{A3} vs.\ \texttt{A4}).
However, it appeared that our model benefited less from beam search decoding than the baseline model,
likely because the LLM’s confidence in its predictions restricted the search space.
Nonetheless, even with $B=1$, our model still surpassed the baseline with $B=20$ (\texttt{A2} vs.\ \texttt{A3}).
This shows the potential to reduce the computational burden required for running the LLM during inference,
as observed by the real-time factor (RTF) measured using a single A100 GPU.

\begin{table}[t]
    \centering
    \vspace{-0.1cm}
    \caption{Ablation studies on LS-100, validating effectiveness of LLM and prompt in our proposed model.}
    \label{tb:ablation_study}
    \vspace{-0.25cm}
    \renewcommand{\arraystretch}{0.95}
    \scalebox{0.95}{
    \begin{tabular}{lcccc}
        \toprule
        & \multicolumn{4}{c}{\textbf{Dev WER}} \\
        \cmidrule(l{0.3em}r{0.3em}){2-5}
        & \multicolumn{2}{c}{$B=1$} & \multicolumn{2}{c}{$B=20$} \\
        \cmidrule(l{0.3em}r{0.3em}){2-3}\cmidrule(l{0.3em}r{0.3em}){4-5}
        \texttt{ID}\ \ \textbf{Model} & clean & other & clean & other \\
        \midrule
        \texttt{-{}-}\ \ Joint CTC/Attn. (\texttt{A1}, \texttt{A3}) & 9.9 & 20.6 & 7.2 & 17.5 \\
        \cdashlinelr{1-5}
        \texttt{-{}-}\ \ + LLM-Guided Dec. (\texttt{A2}, \texttt{A4}) & \textbf{6.7} & \textbf{17.5} & \textbf{6.2} & \textbf{16.5} \\
        \texttt{B1}\ \ \phantom{+} w/o LLM & 9.3 & 20.7 & 7.1 & 17.8 \\
        \texttt{B2}\ \ \phantom{+} w/o Prompt & 9.3 & 19.7 & 6.5 & 16.7 \\
        \texttt{B3}\ \ \phantom{+} w/\phantom{o} Mismatched Task Inst. & 6.8 & 17.8 & 6.5 & 16.8 \\
        \bottomrule
    \end{tabular}
    }
    \vspace{-0.1cm}
\end{table}

\vspace{-0.05cm}
\subsection{Ablation Study}
\vspace{-0.05cm}
\label{ssec:ablation_study}
We conducted several ablation studies for the proposed model to assess the effectiveness of
both the use of the LLM and the prompt.
Table~\ref{tb:ablation_study} presents the results of these ablation studies on LS-100.

\paragraph{Importance of LLM}
We ablated the LLM from the proposed model (\texttt{B1}) by
training the decoder from scratch without Llama2 during Step 2 of the training process in Sec.~\ref{ssec:training}.
Compared to the baseline results, this ablated training yielded only marginal improvement (\texttt{A1}, \texttt{A3} vs. \texttt{B1}).
The proposed model, in contrast, achieved significantly better results thanks to the utilization of the LLM (\texttt{A2}, \texttt{A4} vs.\ \texttt{B1}).

\paragraph{Influence of Prompt}
First, we trained the LLM-guided decoder by removing the prompt from the LLM input (\texttt{B2}),
setting $W^{\mathsf{ins}} = \varnothing$ and $\tilde{W} = \varnothing$ in Eq.~\eqref{eq:llama2}.
While this modification resulted in modest gains compared to the baseline model (\texttt{A1}, \texttt{A3} vs.\ \texttt{B2}),
it adversely affected the performance of the proposed model (\texttt{A2}, \texttt{A4} vs.\ \texttt{B2}).
Next, we trained the LLM-guided decoder with a task instruction that diverged from grammatical error correction (\texttt{B3}),
specifying $W^{\mathsf{ins}}$ to
``\textit{You will be provided with a statement in quotes, and your task is to translate it into Japanese.}''.
While the modified prompt yielded improvements over the baseline (\texttt{A1}, \texttt{A3} vs.\ \texttt{B3}),
the proposed model with the proper prompt demonstrated superior performance (\texttt{A2}, \texttt{A4} vs.\ \texttt{B2}).
The findings from \texttt{B2} and \texttt{B3} suggest that
designing an appropriate prompt is crucial in our model to effectively harness the capabilities of LLMs
for assisting the target speech task.

\begin{table}[t]
    \centering

    \caption{WERs [\%] ($\downarrow$) for joint CTC/attention model using various LLM integration methods on LS-100.}
    \label{tb:lm_integration}
    \vspace{-0.25cm}
    \renewcommand{\arraystretch}{0.95}
    \scalebox{0.95}{
    \begin{tabular}{lcccc}
        \toprule
        & \multicolumn{2}{c}{\textbf{Dev WER}} & \multicolumn{2}{c}{\textbf{Test WER}} \\
        \cmidrule(l{0.3em}r{0.3em}){2-3}\cmidrule(l{0.3em}r{0.3em}){4-5}
        \textbf{Integration Method} & clean & other & clean & other \\
        \midrule
        Joint CTC/Attn. (\texttt{A3}) & 7.2 & 17.5 & 7.5 & 18.0 \\
        + Shallow Fusion & 6.4 & 17.1 & 7.0 & 17.5 \\
        + Rescoring & \textbf{6.1} & \textbf{15.6} & \textbf{6.4} & \textbf{16.1} \\
        + Zero-Shot GEC & 13.8 & 22.8 & 13.4 & 23.3 \\
        \cdashlinelr{1-5}
        + LLM-Guided Dec. (\texttt{A4}) & 6.2 & 16.5 & 6.7 & 16.9 \\
        ++ Rescoring & \textbf{5.3} & \textbf{14.8} & \textbf{5.8} & \textbf{15.1} \\
        \bottomrule
    \end{tabular}
    }
    \vspace{-0.1cm}
\end{table}

\vspace{-0.05cm}
\subsection{Comparison with Conventional LM Integration Methods}
\vspace{-0.05cm}
\label{ssec:lm_integration}

Table~\ref{tb:lm_integration} lists WERs for LS-100,
reporting the performance of the baseline model (\texttt{A3}) when decoded using the LLM (i.e., Llama2).
Here, the proposed approach (\texttt{A4}) is compared with
shallow fusion, rescoring, and zero-shot grammatical error correction (GEC) methods.
Zero-shot GEC was performed by directly evaluating the response generated by the LLM,
given the instruction (detailed in Sec.~\ref{ssec:experimental_setup}) to improve the CTC decoding results (i.e., \texttt{A0}).
Both shallow fusion and rescoring resulted in notable performance improvements over the baseline.
Zero-shot GEC, in contrast, appeared to be challenging, as the LLM tended to produce hallucinations or overcorrections.
These issues led to generating words not present in the speech input,
indicating a need for a dedicated mechanism to align the LLM outputs with the speech information.
While rescoring delivered promising results,
the proposed model demonstrated compatibility with rescoring to achieve the best overall performance.

\vspace{-0.05cm}
\section{Conclusion}
\vspace{-0.05cm}
We proposed to use an instruction-tuned LLM for guiding the text generation process in end-to-end ASR.
By integrating the LLM as a frontend feature extractor for the decoder
and utilizing a precise prompting strategy in conjunction with CTC decoding,
our model effectively utilized linguistic features from the LLM to improve ASR performance.
One inherent limitation of the proposed model is its high computational cost,
as evidenced by RTFs reported in Table~\ref{tb:main_results}.
Future research should explore the use of lightweight, efficient LLMs,
which are actively studied in NLP field~\cite{zhu2023survey}.

\newpage

\bibliographystyle{IEEEtran}
\bibliography{refs_short}

\end{document}